\documentclass[letterpaper, 10 pt, conference]{ieeeconf}
\IEEEoverridecommandlockouts
\usepackage{calc}
\usepackage{mathtools}
\usepackage{amssymb}

\usepackage{amsthm}
\allowdisplaybreaks
\makeatletter
\def\th@definition{
  \thm@notefont{\fontseries\mddefault\upshape}
  \thm@headpunct{}
  \itshape
}
\makeatother
\usepackage{algorithmic}
\usepackage{graphicx}
\usepackage{textcomp}
\usepackage{xcolor}
\usepackage{outlines}
\usepackage{relsize}
\usepackage{mathrsfs}
\usepackage[hidelinks]{hyperref}
\usepackage[capitalize]{cleveref}

\creflabelformat{equation}{#2\textup{#1}#3}

\theoremstyle{definition}
\newtheorem{theorem}{Theorem}

\newtheorem{lemma}{Lemma}
\newtheorem{remark}{Remark}
\newtheorem{definition}{Definition}
\newtheorem{assumption}{Assumption}

\newtheoremstyle{named}{}{}{\itshape}{}{\bfseries}{}{.5em}{\thmnote{#3}}
\theoremstyle{named}

\usepackage{amsmath}

\DeclareMathOperator*{\argmin}{arg\,min}
    
\begin{document}

\title{Incremental Composition of Learned Control Barrier Functions in Unknown Environments
}

\author{Paul Lutkus, Deepika Anantharaman, Stephen Tu, and Lars Lindemann 
\thanks{Paul Lutkus and Lars Lindemann are with the Thomas Lord Department of Computer Science, University of Southern California, Los Angeles, CA, USA. Deepika Anantharaman performed this work while affiliated with the Department of Electrical and Systems Engineering, University of Pennsylvania, Philadelphia, PA, USA.
Stephen Tu is with the Ming Hsieh Department of Electrical and Computer Engineering,
University of Southern California.
	Corresponding E-mail: {\tt\small lutkus@usc.edu}.   
}
}

\maketitle

\begin{abstract}
We consider the problem of safely exploring a static and unknown environment while learning valid control barrier functions (CBFs) from sensor data. Existing works either assume known environments, target specific dynamics models, or use a-priori valid CBFs, and thus provide limited safety guarantees for general control-affine systems during exploration. We present a method for safely exploring 
by incrementally composing a global CBF from local CBFs. The challenge here is that local CBFs may not have well-defined end behavior outside their training domain, i.e. local CBFs may be positive (indicating safety) in regions where no training data is available. We show that well-defined end behavior can be obtained when local CBFs are parameterized by compactly-supported radial basis functions. For learning local CBFs, we collect sensor data, e.g. LiDAR capturing obstacles in the environment, and augment it with simulated data from a safe oracle controller. Our work complements recent efforts to learn CBFs from safe demonstrations, where learned safe sets are limited to their training domains, by demonstrating how to grow the safe set over time as more data becomes available. 
We evaluate our approach on two simulated systems, where our method successfully explores an unknown environment while maintaining safety throughout the entire execution.
\end{abstract}

\section{Introduction}

A key characteristic of a truly autonomous agent is the ability to safely explore an unknown environment by iteratively gathering data about its surroundings, in a safe manner, and then replanning based on its new belief. Consider the example of a mobile search and rescue robot placed in an abandoned warehouse. Upon entering its environment, the robot has only an incomplete understanding of the obstacles in this previously unseen warehouse, and hence must explore the environment, using its 
onboard sensors to gather information about the surrounding obstacles. If this exploration is not done in a safe manner, the robot could 
enter situations where it becomes damaged and can no longer complete its mission.
For real-world, mission-critical deployments, handling this inherent tension between safety and liveliness in a mathematically rigorous way is of paramount importance.


In this work, we attack the safe exploration problem by incrementally composing learned
control barrier functions (CBFs). 
In particular, we extend prior work in offline CBF learning~\cite{robey2020learning,lindemann2024learning,zhao2021learning,sun2021learning} to the online setting by incrementally learning \emph{local} CBFs that individually ensure that the system acts safely with respect to its immediate surroundings. 
Learning local CBFs has considerable computational advantages, especially for long-running agents, since the complexity of each learning problem scales only with the amount of new data collected, and not with cumulative data. 
We compose these local CBFs into a single valid global CBF via a pointwise maximum, using recent progress on non-smooth CBFs~\cite{glotfelter2018boolean}. 
In order to ensure that our CBFs can be 
composed in such a manner, 
we utilize compactly supported basis functions~\cite{wendland1995piecewise}
which ensure negative behavior off the support of the training data.
Together, these ingredients allow us to 
expand the set of safe states that an autonomous system is able to operate in, as more data becomes available, 
while provably remaining safe throughout
exploration 
by inheriting the rigorous safety guarantees of CBFs.

\textbf{Contributions:} (1) We learn local CBFs $h_i$ such that
each $h_i$ is valid on a domain $D_i\subset\mathbb{R}^{d_x}$ \emph{(locally-valid)} and is negative on all $\mathbb{R}^{d_x}\setminus D_i$ \emph{(globally well-defined)},
%
by leveraging compactly supported basis functions to obtain negative end-behavior \cite{wendland1995piecewise}. (2) We provide conditions on the basis function parameters and CBF-synthesis QP to certify that valid local CBFs compose, via pointwise maximum, to a valid, non-smooth global CBF. (3) We efficiently compose local safety computations obtained online via an oracle (e.g. Hamilton-Jacobi (HJ) reachability, model predictive controller (MPC)) through the use of local CBFs.
(4) We provide simulations that illustrate the correctness and applicability of our method. 




\textbf{Related Work:} 
For a known dynamical system and
environment, 
there is a rich body of literature
on safe control synthesis
via both HJ reachability \cite{bansal2017hamilton} and control barrier functions (CBF) \cite{ames2019control}, which are closely related~\cite{choi2021robust}. 
Most similar to our work is \cite{bajcsy2019efficient}, which iteratively computes a HJ-reachability-based least-restrictive controller in contrast to our CBF-based safety filter; we compare the computation time of our method with this work in \cref{sec:experiments}.
Furthermore, generalizations
of both HJ reachability and CBFs
that handle complex system types---including multi-agent and hybrid systems---through the framework of composition
have been recently explored~\cite{glotfelter2018boolean,lindemann2018control,black2023consolidated,yang2024safe}. 


In this work, we primarily focus on CBFs, motivated by their practical use in designing safe control laws via convex programming~\cite{ames2019control}.
However, directly applying CBFs to the problem of safe exploration is difficult for a multitude of reasons.
First, even when both the system and environment are known beforehand, deriving a valid CBF from first principles is a challenging problem. Recent work has focused instead on using safe expert-demonstration trajectories to learn a CBF directly from data~\cite{robey2020learning,zhao2021learning,sun2021learning}.
While learning-based approaches significantly expand the applicability of CBFs,
we cannot na{\"{i}}vely apply these techniques to safe exploration since we do not have a full description of the environment, nor a complete set of safe demonstration trajectories that covers the entire safe space. CBFs have also been applied in safe reinforcement learning, where the main challenge is that the system dynamics are unknown unlike in our work ~\cite{gu2022review}. We view our work as complementary as we expect our methods to compose well with online dynamics estimation.

\par
Recent works have addressed the issue of safe exploration with CBFs  
for various special cases.
The work \cite{gao2023online}
considers online tuning of parameterized CBFs specialized to single-integrator, multi-agent systems.
In \cite{luo2022sample}, the authors show how to ensure simultaneous safety and exploration when learning a general
stochastic, discrete-time system, assuming that
a ground truth CBF is available a-priori.
Furthermore, the works \cite{li2021instantaneous,srinivasan2020synthesis,long2021learning,lafmejani2022nmpc,zhang2024online,safari2024time}
consider the use of sensor data (particularly LiDAR data) to obtain CBFs online. However, these works either consider specialized dynamics models~\cite{lafmejani2022nmpc, li2021instantaneous}, 
require retraining on the entire demonstration dataset whenever new measurements are gathered~\cite{srinivasan2020synthesis}, or restrict CBFs to e.g. elliptical sub-level sets or signed distance functions~\cite{long2021learning,safari2024time,zhang2024online}.
The latter restriction is not guaranteed to satisfy
dynamic feasibility---especially under input constraints or underactuation---and hence cannot guarantee safety in general; we illustrate this issue with
a Dubins car example in \Cref{sec:experiments}.
Finally, we note that in the special case of 
linear dynamics with polytopic constraints, exact controlled-invariant sets can be computed~\cite{anevlavis2023controlled}.

\section{Background}




Consider the control-affine system $\dot{x}=f(x)+g(x)u$ where $x\in\mathbb{R}^{d_x}$ and $u\in\mathbb{R}^{d_u}$ are states and control inputs, respectively. We wish to ensure that the system's trajectory $x(t)$ remains within a  compact set $S\subseteq \mathbb{R}^{d_x}$ for all times $t\ge 0$. To accomplish this, we use control barrier functions (CBFs)~\cite{ames2019control}, which
certify the existence of control inputs guaranteeing 
\emph{forward invariance} of the set $S$.
\begin{definition}[Control Barrier Function]\label{sec:background}
Given a compact set $S\subseteq \mathbb{R}^{d_x}$, a set of  control inputs $U\subseteq \mathbb{R}^{d_u}$, and a locally Lipschitz continuous extended class-$\mathcal{K}$ function $\alpha:\mathbb{R}\to\mathbb{R}$\footnote{An extended class-$\mathcal{K}$ function $\alpha$ is strictly increasing and s.t. $\alpha(0)=0$.}, a continuously differentiable function $h:D\rightarrow\mathbb{R}$ is said to be a control barrier function on an open set $D\supseteq S$ if:
\begin{subequations}
\begin{align}
h(x)&>0\ \forall x\in\mathrm{int}(S)\label{eq_cbf_positivity_property}\\
h(x)&<0\ \forall x\in D \setminus S\label{eq_cbf_negativity_property}\\
h(x)&=0\ \forall x\in\mathrm{bd}(S)\label{eq_cbf_zero_property}\\
\ \sup_{u\in U}\langle\nabla h(x)&,\ f(x)+g(x)u\rangle\geq-\alpha(h(x))\ \forall x\in D.\label{eq_cbf_derivative_property}
\end{align}
\end{subequations}
\label{def_cbf}
\vspace*{-4.5mm}
\end{definition}
%
 Given a reference control signal $u_r:\mathbb{R}_{\ge 0}\to U$, i.e. an open-loop control signal that we would like to match, we can use the CBF $h$ as a safety filter to keep the trajectory $x(t)$ within the set $S$ while closely following $u_r$. 
\begin{lemma}[Safety Filter \cite{ames2019control}]\label{def_cbf_qp}
Given a CBF $h:D\rightarrow\mathbb{R}$ and a reference control signal $u_r:\mathbb{R}_{\ge 0}\to U$, we compute a control signal $u^*:\mathbb{R}_{\ge 0}\to U$ that renders the set $S$ forward invariant by solving the convex quadratic program:
\begin{align}
\begin{split}\label{eq_cbf_qp}
u^*(t)&=\argmin_{u \in U} \quad\|u-u_r(t)\|_2^2\\
\mathrm{s.t.}&\left\langle\nabla h(x(t)), f(x(t))+g(x(t))u\right\rangle\geq-\alpha(h(x(t))).
\end{split}
\end{align}
\end{lemma}
In the remainder, we will refer to the quadratic program in \cref{eq_cbf_qp} as the CBF-QP.

\begin{figure}[t]
\centerline{\includegraphics[width=0.35\textwidth]{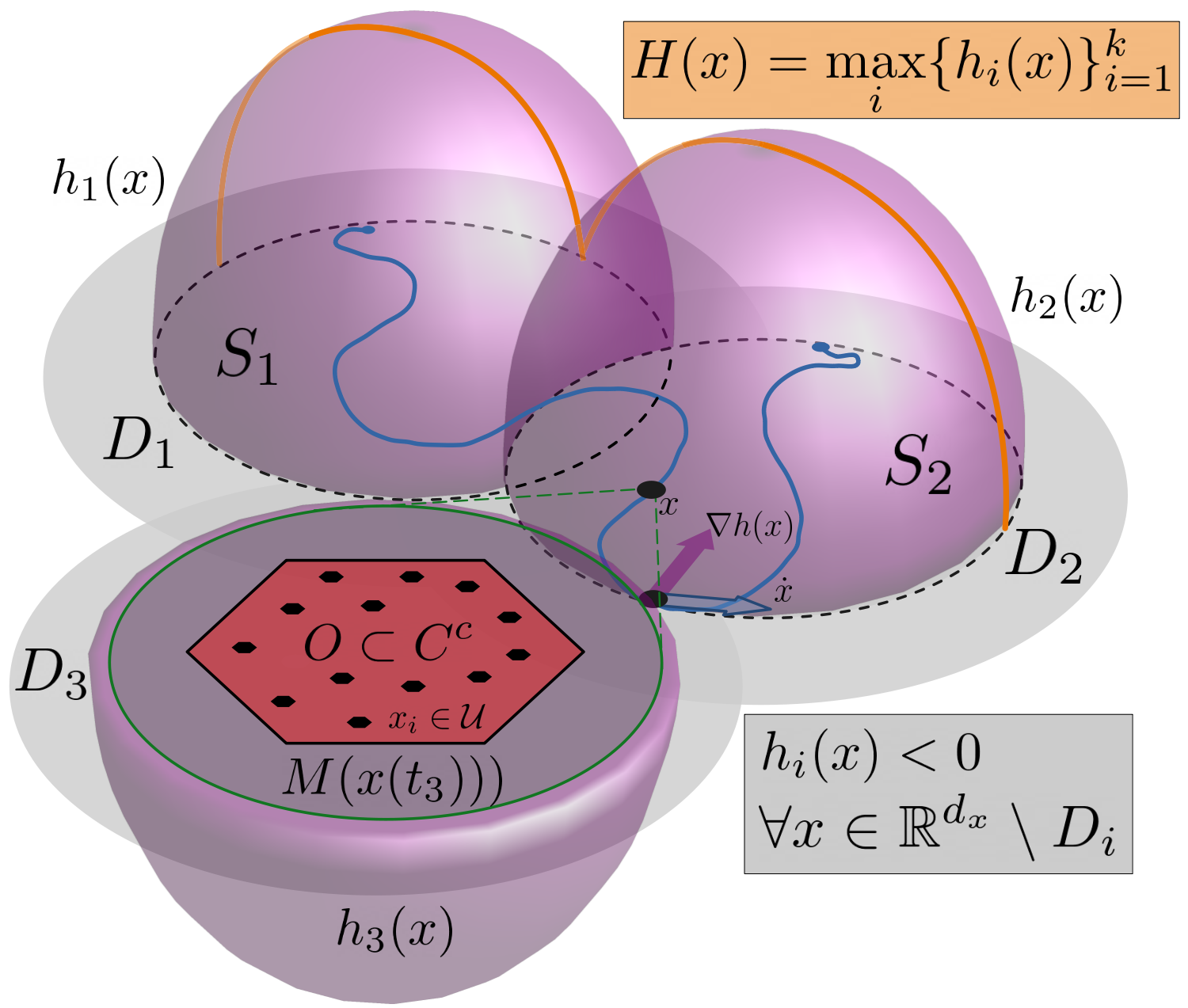}}
\vspace*{-3mm}
\caption{Scan $M(x(t_3))$ detects red hexagonal obstacle $O$, learns negative $h_3$ with no invariant set $S$, system remains in $\max$ CBF.}
\label{figcbf}
\vspace*{-6mm}
\end{figure}

\subsection{Learning CBFs from Data}\label{learning_cbfs}

Constructing valid CBFs is generally challenging and has motivated learning-based approaches. We briefly review our prior work in \cite{robey2020learning,lindemann2024learning} on learning CBFs from demonstrations, which serves as a basis for this paper. 
%
%
Here, an expert dataset $\mathcal{D}:=\{(x_i, u_i)\}_{i=1}^N$ is given where $N$ state-action pairs demonstrating safe system behavior. From the dataset $\mathcal{D}$,
a CBF of the form
$h(x):=\sum_{j=1}^{J}\theta_j\phi_j(x)=\theta^T\phi(x)$ is learned by solving a quadratic program (QP) where $\phi(x):\mathbb{R}^{d_x}\rightarrow\mathbb{R}^J$ are $J$ pre-defined basis functions and $\theta\in\mathbb{R}^J$ are the decision variables.
%
%
This QP is formed by constructing three 
distinct datasets using information from $\mathcal{D}$. The buffer dataset $\mathcal{B}$
obtained via boundary point detection algorithms (see \cite{lindemann2024learning} for details),
the safe dataset $\mathcal{S}:=\mathcal{D}\setminus\mathcal{B}$, 
and the unsafe dataset $\mathcal{U}$
obtained by sampling or again applying boundary point detection, together with detected obstacles.

\begin{definition}[Learning-CBF QP \cite{robey2020learning}]\label{def_learning_cbfs_qp}
Given the datasets $\mathcal{S}$, $\mathcal{B}$, and $\mathcal{U}$, the hyperparameters $\gamma_\text{safe}, \gamma_\text{unsafe},  \gamma_\text{dyn}>0, b>0$, a locally Lipschitz continuous extended class-$\mathcal{K}$ function $\alpha:\mathbb{R}\to\mathbb{R}$, and a basis function $\phi:\mathbb{R}^{d_x}\rightarrow\mathbb{R}^J$ with bounded Lipschitz constant, we learn a candidate CBF $h(x):=\theta^T\phi(x)-b$ by solving  the convex quadratic program:
\begin{subequations}\label{eq:learning_CBF_QP}
\begin{alignat}{2}
\min_{\theta\in\mathbb{R}^J}\quad&\|\theta\|_2^2\\
\mathrm{s.t.}\quad&\theta^T\phi(x_i)-b\geq\hspace{\widthof{$-$}}\gamma_\text{safe}&&\mathllap{\forall x_i\in\mathcal{S}\quad\quad}\label{eq_learning_cbfs_positivity_constraint}\\
&\theta^T\phi(x_i)-b\leq-\gamma_\text{unsafe}&&\mathllap{\forall x_i\in\mathcal{U}\quad\quad}\label{eq_learning_cbfs_negativity_constraint}\\
&\left\langle D_x\phi(x_i)^T\theta,\ f(x_i)+g(x_i)u_i\right\rangle+\alpha (h(x_i))\geq\gamma_\text{dyn}\label{eq_learning_cbfs_derivative_constraint}\nonumber\\
& &&\mathllap{\forall (x_i, u_i)\in\mathcal{S}\cup\mathcal{B}\quad\quad}
\end{alignat}
\end{subequations}
where $D_x\phi(x)$ denotes the Jacobian of $\phi(x)$. 
\end{definition}

We refer to \eqref{eq:learning_CBF_QP} as the Learning-CBF QP, and remark that the candidate CBF obtained by solving \eqref{eq:learning_CBF_QP} is valid when conditions on the density of points in $\mathcal{S}$, $\mathcal{B}$, and $\mathcal{U}$ and on the Lipschitz constants of $h$, $f$, and $g$ are satisfied~\cite{robey2020learning}. 
Here, we use the Learning-CBF QP to learn local CBFs $h_i$ that are locally-valid on a domain $D_i$. 
In \cite{robey2020learning}, the basis functions $\phi(x)$ are chosen to be random Fourier features (RFF)~\cite{rahimi2007random}, which take the form $\phi(x)=\cos(Wx+b)$ for randomly sampled $W\in\mathbb{R}^{J\times d_x}, b\in\mathbb{R}^J$, where $\cos(\cdot)$ is applied element-wise.
Unfortunately, the oscillatory end-behavior of RFF prevents composition of learned CBFs via the $\max$-operator (cf.~\cref{figcbf}), since the local CBFs $h_i$ may oscillate to positive outside of their domain $D_i$ i.e., for a radius $r$ and point $x$, we can find $x'$ with $\|x-x'\|_2>r$ such that $\phi_j(x')>0$. Obtaining negative end-behavior and thus global well-definedness motivates using compactly supported RBFs. 

\subsection{Compactly Supported Radial Basis Functions}\label{csrbf}
Radial basis functions (RBFs) are scalar functions with radial symmetry about a center $z\in\mathbb{R}^{d_x}$, i.e. ${\phi}(x,z):=\bar{\phi}(\|x-z\|_2)$
for a $\bar{\phi}:\mathbb{R}\to\mathbb{R}$. For a set of $J$ centers $Z:=(z_1,\hdots,z_J)$, we combine RBFs linearly
as $h(x) := \theta^T \phi(x, Z)-b$, where
$\phi(x, Z) := ({\phi}(x, z_1),\hdots, {\phi}(x, z_J))$. 
%
We consider a specific family of RBFs, referred to as
compactly supported radial basis functions (CS-RBFs).
CS-RBFs are zero outside of a radius $s$ around $z$, which ensures that $h_i$ is negative outside of its validity region $D_i$ for $b>0$, and thus globally well-defined.
%
When using CS-RBFs to learn CBFs, we require continuous differentiability, and so we use Wendland's functions, which are compactly supported, polynomial basis functions that are minimal in their polynomial degree \cite{wendland1995piecewise}. 
An example of a Wendland's function is $\bar{\phi}(r) := \max(0, 1-r)^4(1+4r)/20$ for $d_x \leq 3$.


\section{Safe Exploration via Incremental Composition of CBFs}
\label{sec:our_method}


We now describe our approach to safe exploration via incrementally composing local CBFs in an unknown environment. 
Let $C\subseteq\mathbb{R}^{d_x}$ denote the a-priori unknown geometric safe region,
i.e. the constraint set that describes the obstacle-free space. Denote $q:=(x^1,\hdots,x^{d_q})$ for $d_q \leq d_x$ as the observable coordinates from a subset of the full coordinates $x:=(x^1,\hdots,x^{d_x})$ for which safety can be determined by a measurement map $M:\mathbb{R}^{d_x}\to\text{proj}_q(2^C)$ that takes the state $x$ to a subset of the geometric safe region, projected\footnote{$\text{proj}_q(S)\!=\!\{q\in\mathbb{R}^{d_q}\!:\exists(q,x^{d_q+1},\hdots,x^{d_x})\!\in\!S\}$,\,\,$2^S\!=\!\{X\!\!:\!\!X\subseteq S\}$} onto the coordinates $q$. $M(x)$ represents local information obtainable about $C$ from state $x$; for example, spatial measurements collected from onboard sensors, e.g. LiDAR. For a sensor that detects all obstacles $P\subseteq\mathbb{R}^{d_q}$ in a radius $r$ around $x$, $M(x)=P^c\cap\mathcal{B}_r(x)$, where $\mathcal{B}_r(x)$ is a ball of radius $r$ centered at $x$ and $P^c$ is the complement of $P$.

Our goal is to 
collect measurements $M(x(t))$ 
and build a composite CBF $H(x)$ that defines a safe set $S:=\{x \in \mathbb{R}^{d_x} : H(x) \geq 0\}$ that covers as much of $C$ as possible, while respecting $x(t) \in C$ for all times.
We solve this safe exploration problem by 
combining local CBFs $h_i$ as follows.
At time $t=0$, we take a measurement $M(x(0))$. We sample $M(x(0))$ (and $M(x(t_k))$ in future iterations) to generate a dataset $\mathcal{D}$, from which $\mathcal{S}, \mathcal{B}$, and $\mathcal{U}$ are constructed as discussed in \cref{learning_cbfs}; we discuss how to obtain control inputs for this dataset in \cref{def_safety_oracle}. Using the Learning-CBF QP (cf.~\Cref{def_learning_cbfs_qp}), we learn an initial CBF $h_0$  which is valid on some set $D_0\subseteq \mathbb{R}^{d_x}$, and renders a subset $S_0 := \{ x \in D_0 : h_0(x) \geq 0 \}\subseteq C$ forward invariant.\footnote{Here, $D_0$ is the set on which the constraints in~\Cref{def_learning_cbfs_qp} are satisfied.} 
In addition, we require that $S_0$ agrees with the measurement, i.e. $\text{proj}_q(S_0)\subseteq M(x_0)$.
We then repeat this process. Let $k$ denote the number of measurements taken so far (starting at $k=1$), and let $t_k$ denote the time the $k$-th measurement is taken at.
%
We summarize the procedure as follows:
\begin{enumerate}
    \item Let $H_k(x) := \max_{i \in \{0, \dots, k-1\}} h_i(x)$
    denote the CBF that renders $\cup_{i \in \{0, \dots, k-1\}} S_i$ forward invariant.
    \item Use $H_k$ to approach the boundary of the current  invariant set $\cup_{i \in \{0, \dots, k-1\}} S_i$ (cf.~\Cref{remark_cbf_descent}).
    \item Take a new measurement $M_k=M(x(t_k))$, and learn a CBF $h_k$ which is valid on some set $D_k\subseteq \mathbb{R}^{d_x}$, renders a subset $S_k := \{ x \in D_k : h_k(x) \geq 0 \}$ forward invariant, and satisfies $\text{proj}_q(S_k)\subseteq M_k$.
    \item Increment $k \gets k + 1$.
\end{enumerate}
%
The key insight of our procedure is that
by ensuring that the learned invariant set $S_k$ satisfies $\text{proj}_q(S_k)\subseteq M_k=M(x(t_k))$, we keep the system safe for all time.

\begin{remark}\label{remark_cbf_descent}
In order to approach the boundary of the current invariant set $\cup_{i \in \{0, \dots, k-1\}} S_i$, one can choose a point $x^-$ such that $H_k(x^-)<0$, and construct an MPC that steers $x(t)$ towards $x^-$. The control signal $u_r(t)$ generated by this MPC can then be filtered by  the CBF-QP in \cref{eq_cbf_qp} to obtain a signal $u(t)$ that safely approaches the boundary of $\cup_{i \in \{0, \dots, k-1\}} S_i$.
\end{remark}

Learning CBFs in an incremental manner when new measurements $M_k$ become available yields an efficient algorithm where one 
need not recompute CBFs for regions in the state space where no new information about the set $C$ is obtained. However, this incremental CBF construction raises two important questions.
The first,
which we address in \Cref{sec:expert_demos_online}, is how one obtains the necessary expert dataset in Step 3) in order to apply the Learning-CBF QP while satisfying $\text{proj}_q(S_k)\subseteq M_k$.
The second,
addressed in 
\Cref{sec:composing_local_CBFs}, is
what conditions are needed to ensure that the $\max$ composition in Step 1) yields a valid global CBF.

\subsection{Obtaining Expert Demonstrations Online}
\label{sec:expert_demos_online}

\begin{figure*}{}
\centering
\makebox[\textwidth][c]{\includegraphics[width=0.98\textwidth]{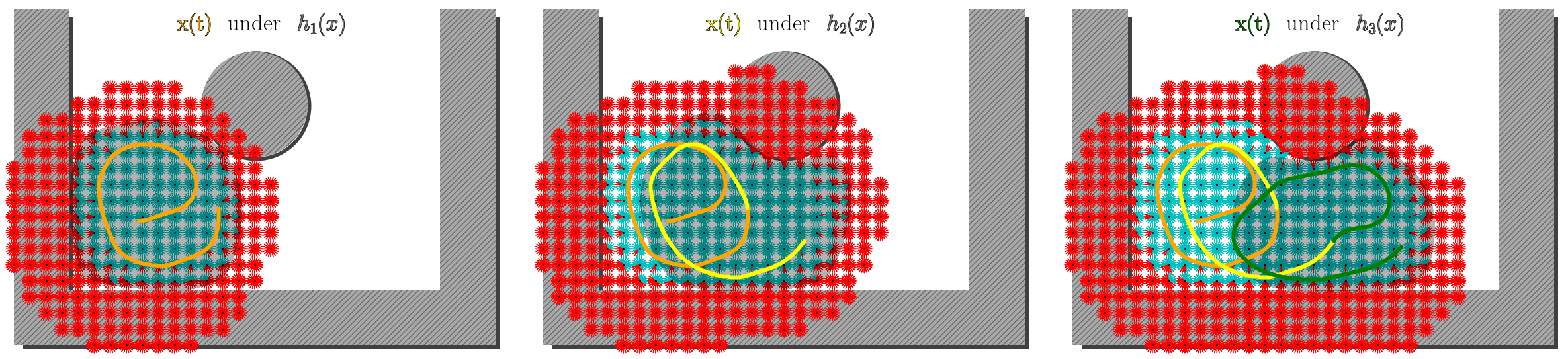}}
\vspace*{-3mm}
\caption{\emph{Left-to-Right}: Incremental learning of global Dubins car CBF, while safely exploring unknown environment under input constraints. LiDAR scans are taken where the trajectory changes color. Showcasing the correctness of local CBFs, each trajectory is constrained by the last-learned local CBF, whose zero-level set is shaded. At each $(q^1,q^2)$, red arcs denote unsafe angles $h(q^1,q^2,\theta)<0$. Blue arcs denote $h(q^1,q^2,\theta)\geq0$.}
\label{fig_bicycle_exploration1}
\vspace*{-5mm}
\end{figure*}

In order to apply the Learning-CBF QP described in \Cref{learning_cbfs} 
we need  access to a dataset of expert demonstrations containing safe state-action pairs (cf.~\Cref{def_learning_cbfs_qp}).
%
This may not be possible during exploration. Suppose our measurement map is implemented via a LiDAR sensor. While LiDAR provides us with a point cloud of obstacles, it does not provide expert controls that avoid the obstacles. We provide two remedies for the lack of safe inputs. Our first solution is based on the dual norm. Suppose that the input set $U$ is a norm ball of the form $\|u\| \leq u_{\max}$.
Then, the CBF derivative constraint $\sup_{u\in U}\langle\nabla h(x),\ f(x)+g(x)u\rangle\geq-\alpha(h(x))$ and hence the constraint \cref{eq_learning_cbfs_derivative_constraint} in the Learning-CBF QP can be rewritten in terms of the dual norm
$\| u \|^\star = \sup_{x \in \mathbb{R}^{d_n}} \{ \langle x, u \rangle : \| x \| \leq 1 \}$ as
follows~\cite{lindemann2021learning}: $\langle\nabla h(x),f(x)\rangle+ u_{\max} \|g(x)^T\nabla h(x)\|^\star+\alpha(h(x))\geq\gamma_{\text{dyn}}$.
The above equation does not require safe inputs $u$, but results in a nonconvex Learning-CBF QP.
A deeper problem is that sensors may not fully characterize the safe state (i.e. $d_q < d_x$). LiDAR on a car may provide safe \& unsafe positions, but not safe steering angles and velocities, which are critical to avoid states that inevitably lead to collision. To address the limitations of the dual norm, our second approach defines the notion of a \emph{safety oracle}, which takes the safe set returned by a sensor as input and returns safe state-action pairs for which there exists a control signal that keeps the system in this safe set for all time.

\begin{definition}[Safety Oracle]\label{def_safety_oracle}
Recall that $q\in\mathbb{R}^{d_q}$ denotes the observed subset of the full coordinates $x\in\mathbb{R}^{d_x}$. For some state $x_s\in\mathbb{R}^{d_x}$ at which a measurement is obtained, assume that the measurement map $M(x_s)$ returns a local safe subset  $C_\mathrm{loc}\subset\text{proj}_q(C)$ contained in a local scan region $R\supseteq C_\mathrm{loc}$ around $x_s$, i.e. $x=(q, x^{d_q+1},\hdots,x^{d_x}) \notin C$ for all $q\in R\cap C_\mathrm{loc}^c$ and $(x^{d_q+1},\hdots,x^{d_x})\in \mathbb{R}^{d_x-d_q}$.
A safety oracle $\mathcal{O}:M(x_s)\mapsto\{(x_i, u_i)\}_{i=1}^{N_s}$ takes 
the observed safe set $M(x_s)=C_\mathrm{loc}$ and returns some safe state-action pairs $\mathcal{D}=\{(x_i,u_i)\}_{i=1}^{N_s}$ for which there exists a control signal $u(t)$ that ensures $\text{proj}_q(x(t))\in C_\mathrm{loc}$  for all times $ t\geq0$ and for all initial conditions $(x(0),u(0))=(x_i,u_i)\in\mathcal{D}$.
\end{definition}


Some examples of safety oracles are Hamilton-Jacobi reachability analysis \cite{bansal2017hamilton}, safe MPC \cite{frasch2013auto}, and obstacle-aware shooting methods \cite{an2024multiple}. Our proposed approach allows us to efficiently use local oracle computations in two ways. First, while oracle computations may only provide safe state-action pairs for a finite set of states,
the learned locally-valid CBF interpolates between the given datapoints (cf.~\Cref{def_learning_cbfs_qp}), yielding safety guarantees over the entire local domain. Secondly, oracle computatations may be costly. If we recomputed the oracle over all previously seen states, the computational cost would grow in an unbounded manner as the agent explores. By computing the oracle over a local region around the agent, the computational cost stays constant during exploration. By synthesizing local safety computations into composable CBFs, we store and combine them in a principled way,
circumventing the need to recalculate the safety of already-certified trajectories. Furthermore, our approach produces a safety filter (cf. \cref{def_cbf_qp}) that can be used alongside any nominal controller, unlike many oracles alone.
The following example illustrates our approach.

\begin{figure}{}
\centerline{\includegraphics[width=0.35\textwidth]{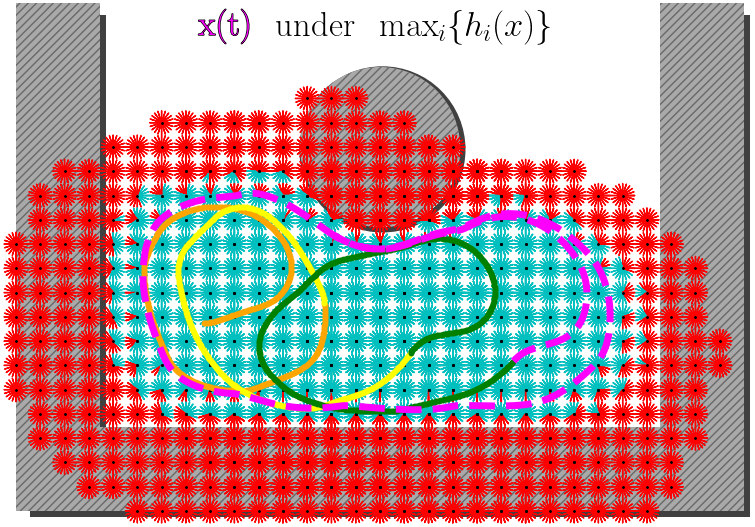}}
\vspace*{-3mm}
\caption{Global CBF for Dubins car. Pink trajectory of system under global CBF interpolates system behavior under local CBFs shown in \cref{fig_bicycle_exploration1}.}
\vspace*{-5mm}
\label{fig_bicycle_exploration2}
\end{figure}

\noindent\emph{Dubins car:}
Consider dynamics $\dot{x}=(\dot{q}^1,\dot{q}^2,\dot{\theta})=(v\cos\theta, v\sin\theta, u)$,
with position $q:=(q^1, q^2)$ and fixed velocity $v$. The system operates in an unknown environment with access to a LiDAR sensor that returns safe and unsafe positions $\mathcal{S}_{q}:=\{q_i\}_{i=1}^{N_s}$ and $\mathcal{U}_{q}:=\{q_i\}_{i=1}^{N_u}$. Critically, the LiDAR does not provide safe or unsafe $\theta$ values, nor safe controls $u$. By joining the unsafe dataset $\mathcal{U}_{q}$ with unsafe positions $\mathcal{Z}_{q}$ sampled from beyond the  convex hull of $\mathcal{S}_{q}$, we can perform Hamilton-Jacobi reachability analysis \cite{bansal2017hamilton} to obtain a value function $V(x)$ that, for each $q\in\mathcal{S}_{q}$, is positive for safe $\theta$'s where there exists a control signal that avoids $\mathcal{U}_{q}\cup\mathcal{Z}_{q}$. 
For each $q\in\mathcal{S}_{q}$ the safe angles are $\{\theta:V(q^1, q^2, \theta)\geq0\}$. For a discretized range of $\theta$'s, the result is a dataset $\mathcal{D}:=\{(x_i, u_i)\}_{i=1}^N$ of safe states and controls, where the gradient of the value function, $\nabla V(x)$, is used to obtain safe controls according to $\pi^*(x)=\arg\max_{u\in U}\langle\nabla V(x),f(x)+g(x)u\rangle$. Synthesizing $\mathcal{D}$ into a locally-valid CBF effectively interpolates the gridded solution to the HJB-PDE, and allows us to combine information from multiple solutions by composing their learned CBFs.

\par

\subsection{Composing Locally-Valid CBFs}
\label{sec:composing_local_CBFs}

In the previous subsection, we generated datasets of safe state-action pairs for obtaining locally-valid CBFs via the Learning-CBF QP.
Consider the task of composing $k$ locally-valid CBFs $h_1, \dots, h_k$. Because $h_i$ is only valid on $D_i$ where its dataset was sampled from, $h_i$ should be negative for $\mathbb{R}^{d_x}\setminus D_i$ in order to be globally well-defined. 
We show that if $h_i(x)<0\ \forall x\in\mathbb{R}^{d_x}\setminus D_i$, then $H(x):=\max_{i \in \{1,\hdots,k\}}h_i(x)$ is a valid nonsmooth CBF on all $\bigcup_{i \in \{1,\hdots,k\}}D_i$.

\par

Let $\phi_s(x,Z):\mathbb{R}^{d_x}\times\mathbb{R}^{d_x\times J}\to\mathbb{R}^J$ be a vector of compactly-supported RBFs that decay to zero in radius $s$. Let each component $[\phi_s(x,Z)]_j$ of $\phi_s(x,Z)$ be a Wendland's function of the form ${\phi}_s(x,z_j):=\bar{\phi}(\|x-z_j\|_2/s)=(1-r_j/s)_+^p f(r_j/s)$ for $r_j:=\|x-z_j\|_2$, where $(\cdot)_+:=\max(0,\cdot)$ and $f(r_j/s)$ is polynomial in $r_j/s$ with order less than $p$ \cite{wendland1995piecewise}. Note that $\bar{\phi}(r_j/s)=0\ \forall r_i\geq s$.
Now assume that $D_i$ is compact and contains all centers $z_j$ of $h_i$. We show that if all centers $z_j$ whose weight $\theta_j$ is positive are at least a distance of $s$ away from $\mathrm{bd}(D_i)$, which can be guaranteed by sampling unsafe points to increase the size of $\mathcal{U}$ in \cref{eq_learning_cbfs_negativity_constraint}, then $h_i(x)<0$ on all $\mathbb{R}^{d_x}\setminus D_i$. It follows that $\bigcup_iS_i$ is forward invariant under controls generated by $H(x)$. The next assumptions precede our main result where $d(z,D)$ denotes the minimum $L_2$ distance from point $z$ to set $D$.

\begin{assumption}[Persistent Feasibility]\label{assumption_feasibility}
$\exists\varepsilon>0$ such that the optimization problem $u^*(t)=\argmin_{u \in U}\|u-u_r(t)\|_2^2$
\begin{align}
\text{subject to}\quad\left\langle\nabla h_i(x), f(x)+g(x)u\right\rangle\geq-\alpha(H(x))&\nonumber\\
&\mathllap{\forall i\in\{i':|h_{i'}(x)-H(x)| \leq \epsilon\}}\label{mod_cbf_qp},
\end{align}
has a bounded solution for all $x\in\bigcup_{i\in\{1,\hdots,k\}}S_i$.
\end{assumption}
\noindent Even if \cref{mod_cbf_qp} is infeasible, safety is never jeopardized, since the agent can simply stay in the current CBF, and reattempt crossing into a neighboring CBF at a later time. Note the assumption of persistent feasibility is also made in \cite{glotfelter2018boolean}.
\begin{assumption}[RBF Decay Buffer]\label{assumption_decay} The centers $Z^{(i)}$ of CBF $h_i$ satisfy the following: For all $z_j^{(i)} \in Z^{(i)}$, $z_j^{(i)} \in D_i$ and $d(z_j^{(i)}, \mathrm{bd}(D_i))\geq s$ if $\theta_j^{(i)} > 0$. 
\end{assumption}
\noindent This assumption is easily satisfied by adding negativity constraints \cref{eq_learning_cbfs_negativity_constraint} for points sampled in a spherical shell of width $s$, with inner radius equal to the outermost center. This a sufficient but not necessary condition, and we find that a shell of width $s'\ll s$ is sufficient in our experiments.

\begin{theorem}[Forward Invariance for $H(x)=\max_i \{h_i(x)\}$]
\label{thm_max_forward_invariance}
Consider $k$ continuously differentiable CBFs $\{h_i(x)\}_{
i=1}^k$ parameterized as $h_i(x):=\phi_s(x,Z^{(i)})^T\theta^{(i)}-b$ with weights $\theta^{(i)}$, centers $Z^{(i)}$, and $b>0$. Let each $h_i(x)$ be locally-valid on a corresponding region from $\{D_i\}_{i=1}^k$. If \cref{assumption_feasibility} and \cref{assumption_decay} hold,
then $H(x)=\max_i \{h_i(x)\}$ is a valid non-smooth CBF that guarantees  forward invariance of the set $\bigcup_{i \in \{1,\hdots,k\}} S_i\subseteq\bigcup_{i \in \{1,\hdots,k\}} D_i$.
\end{theorem}
\noindent\textit{Proof:} 
If $d(z_j^{(i)}, \mathrm{bd}(D_i))\geq s$ for all $z_j^{(i)}$ s.t. $\theta_j^{(i)}>0$, then $b>0\Rightarrow h_i(x)<0$ (strict ineq.) $\forall x\in\mathbb{R}^{d_x}\setminus D_i$ by the compact support property of $\bar{\phi}_s$. By \cite{glotfelter2018boolean} Thm. 3 and convexity of the almost-active gradient, $H(x)$ is a non-smooth CBF yielding forward invariance for the set $\bigcup_{i \in \{1,\hdots,k\}} S_i$. \qed

The set $I_\epsilon(x):=\{i':|h_{i'}(x)-H(x)| \leq \epsilon\}$ is called the ``almost-active'' set in \cite{glotfelter2018boolean}. The QP in \cref{thm_max_forward_invariance} thus
imposes a derivative constraint for each $h_i(x)$ with $i\in I_\epsilon(x)$.


    

\begin{figure}{}
\centering
\includegraphics[width=8cm]{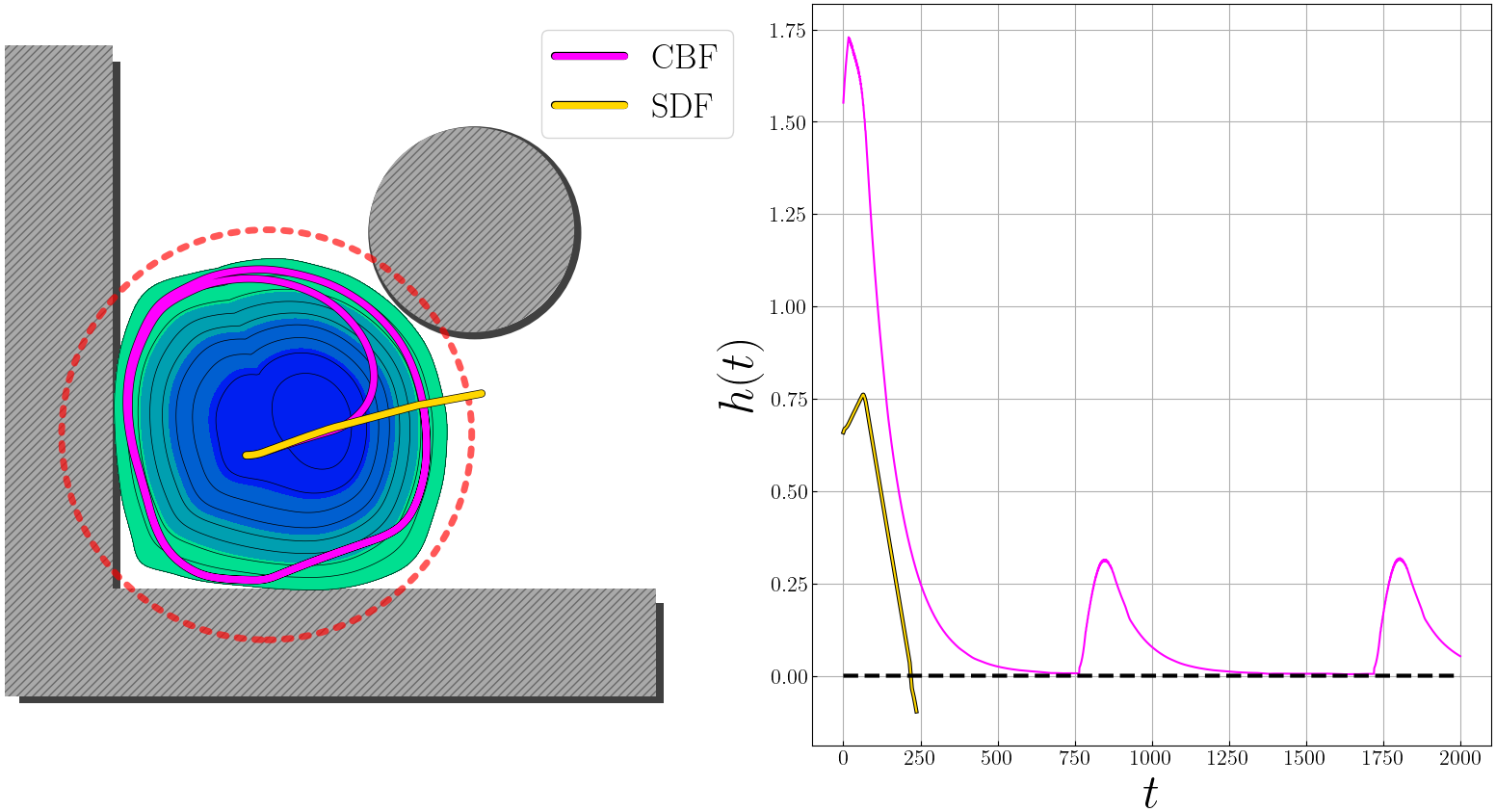}
\vspace*{-3mm}
\caption{CS-RBF barrier (teal contours) constrains Dubins car (pink) to the safe region. Signed distance functions cannot certify fwd. invariance for this system (gold, leaves safe set), even with unbounded actuation.}

\label{fig_invariance}
\vspace*{-5mm}
\end{figure}

\section{Case Studies}
\label{sec:experiments}

To illustrate the validity of our approach, we incrementally learn global CBFs for two systems: a planar, feedback-linearizable system and a fixed-velocity Dubins car.  For both case studies, we choose $s=1$ and obtain CS-RBFs  $\phi_{s=1}(x,z_j):=\max\{0, 1-r_j\}^4(1+4r_j)/20$.  For simplicity, we pick the centers $z_j$ to fall on a grid in $D_i$; more sophisticated arrangements of centers could be chosen, e.g. along safe trajectories provided by the oracle. The code for our case studies can be found at {\small\url{https://github.com/paullutkus/learning-cbfs-online}}.

\noindent\emph{Dubins Car:} The dynamics of the Dubins car are $\dot{x}=(\dot{q}^1,\dot{q}^2,\dot{\theta})=(v\cos\theta, v\sin\theta, u)$ with fixed velocity $v:=0.1$, initial condition $x(0):=(-1.1, -1.1, 0)$, and input constraints $|u|\leq u_\textrm{max}:=0.4$. The environment consists of a central obstacle and four walls (\cref{fig_bicycle_exploration1}). For our idealized LiDAR, we define a circular scan radius of $1.1$ (depicted in red in \cref{fig_invariance}) centered on the agent. Since all states outside of the scan could be obstacles, the local CBF must keep the system within this radius. Our local CBFs $h_i(x)$ and composite CBF $H(x)$ yield invariance while satisfying input constraints (see \cref{fig_bicycle_exploration1,,fig_bicycle_exploration2}). In contrast, candidate CBFs constructed without incorporating
dynamics may not be valid nor provide safety.
To illustrate this, we consider 
signed distance functions (SDFs), which have been used as candidate CBFs in the past, see e.g. \cite{long2021learning}. 
\cref{fig_invariance} shows that an SDF cannot provide safety for this system, even with unbounded actuation,
as the gradient of an SDF to obstacles $l(x)$ is always orthogonal to the vector field of the Dubins car ($\partial l/\partial\theta=0$). While high-order CBFs can resolve this incompatibility, we are able to control the system with a simpler, first-order CBF. We also perform a timing comparison with \cite{bajcsy2019efficient}, assessing the cost of local CBFs when using an HJB oracle. We find a $45\%$ average overhead compared to using local HJB only (2.45s vs 3.55s), suggesting that a composite CBF can be obtained for minimal added cost.

\begin{figure}{}
\centering
\includegraphics[width=9cm]{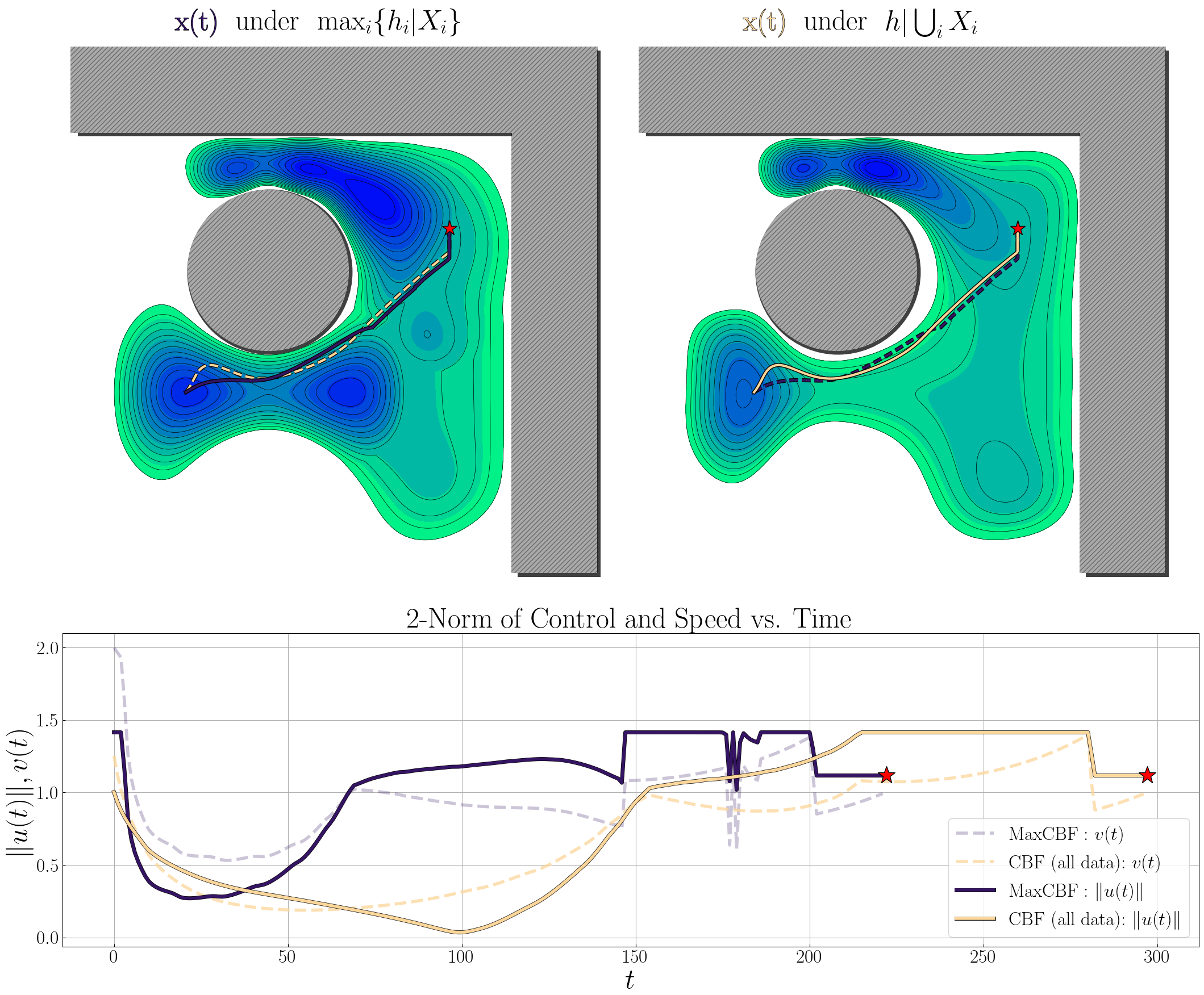}
\vspace*{-7mm}
\caption{$h_i|X_i$ denotes that $h_i$ was obtained from data $X_i$. \emph{Top:} The gradient norms of $\max_i\{h_i|X_i\}$ are visibly larger than those of $h|\bigcup_iX_i$, and the zero-level set of $\max_i\{h_i|X_i\}$ is closer to the central obstacle. \emph{Bottom:} Trajectory under $\max_i\{h_i|X_i\}$ reaches target faster, with larger actuation \& speed while maintaining safety, suggesting 
decreased conservatism.}
\label{fig_planar_comparison}
\vspace*{-5mm}
\end{figure}

\noindent\emph{Planar System:} The dynamics of the planar system are $(\dot{x}_1, \dot{x}_2)=(x_1+(x_1^2+\delta)u_1, x_2+(x_2^2+\delta)u_2)$
with $\delta:=0.33$ and initial condition $x(0)=(0,0)$. We consider the input constraints $|u| \leq u_\textrm{max}=1$. The environment consists of two vertically-aligned circular obstacles and four walls (\cref{fig_planar_comparison}). The  scan radius of the LiDAR sensor is $1$. Here, we compare our incremental CBF $H(x)$ with a CBF trained in a single-shot on all the data that was used to incrementally train $H(x)$. Not only do we match performance with such a single-shot CBF, but our incremental CBF exceeds it in this case (\cref{fig_planar_comparison}). Specifically, the zero-level set is closer to the obstacle, and larger gradients allow the system to approach a target faster. This is not surprising, as the max of continuously-differentiable CBFs is a more expressive function class than single continuously-differentiable CBFs. Furthermore, incremental learning is almost three times faster ($5.2$s total for four local CBFs vs. $14.9$s for one global).

\vspace*{-2mm}
\section{Conclusion}
\vspace*{-1mm}
We presented an online approach for the incremental construction of a  non-smooth CBF via local CBFs, allowing for safe exploration in an unknown, static environment, by relying on a class of compactly-supported basis functions,
whose end behavior guarantees the valid composition of local CBFs. We validated our approach on a feedback-linearizable planar system and a fixed-velocity Dubins car, and compared to existing work. Though we observe good performance with CS-RBFs, we find their Lipschitz constants to be sometimes too large to certify validity for local CBFs via the conditions in $\cite{robey2020learning}$, which are sufficient but not necessary. Future work may benefit from more expressive basis functions
or less-conservative conditions for validity. Another avenue is 
generalization to dynamic and stochastic environments. 



\vspace*{-2mm}
\bibliographystyle{IEEEtran}
\bibliography{ref}

\end{document}